\documentclass[onecolumn, english, reprint, superscriptaddress, breaklinks=true, showkeys, showpacs=false, nofootinbib]{revtex4}

\usepackage[T1]{fontenc}
\usepackage{color}
\usepackage{babel}
\usepackage{float}
\usepackage{amssymb}
\usepackage{listings}
\usepackage{graphicx}
\usepackage{amsmath}
\usepackage{svg}
\usepackage[normalem]{ulem}

\usepackage{tcolorbox}

\newtcolorbox[auto counter, number within=section]{definition}[2][]{colframe=gray!40, colback=white, coltitle=black, fonttitle=\bfseries, title=Definition 1}

\usepackage[unicode=true,pdfusetitle, bookmarks=true,bookmarksnumbered=false,bookmarksopen=false, breaklinks=true,pdfborder={0 0 0},backref=false,colorlinks=true]{hyperref}
\usepackage{breakurl}
\definecolor{Code}{rgb}{0,0,0}
\definecolor{Decorators}{rgb}{0.5,0.5,0.5}
\definecolor{Numbers}{rgb}{0.5,0,0}
\definecolor{MatchingBrackets}{rgb}{0.25,0.5,0.5}
\definecolor{Keywords}{rgb}{0,0,1}
\definecolor{self}{rgb}{0,0,0}
\definecolor{Strings}{rgb}{0,0.63,0}
\definecolor{Comments}{rgb}{0,0.63,1}
\definecolor{Backquotes}{rgb}{0,0,0}
\definecolor{Classname}{rgb}{0,0,0}
\definecolor{FunctionName}{rgb}{0,0,0}
\definecolor{Operators}{rgb}{0,0,0}
\definecolor{Background}{rgb}{0.98,0.98,0.98}
\lstdefinelanguage{Python}{
numbers=left,
numberstyle=\footnotesize,
numbersep=1em,
xleftmargin=1em,
framextopmargin=2em,
framexbottommargin=2em,
showspaces=false,
showtabs=false,
showstringspaces=false,
frame=l,
tabsize=4,
basicstyle=\ttfamily\small\setstretch{1},
backgroundcolor=\color{Background},
commentstyle=\color{Comments}\slshape,
stringstyle=\color{Strings},
morecomment=[s][\color{Strings}]{"""}{"""},
morecomment=[s][\color{Strings}]{'''}{'''},
morekeywords={import,from,class,def,for,while,if,is,in,elif,else,not,and,or,print,break,continue,return,True,False,None,access,as,,del,except,exec,finally,global,import,lambda,pass,print,raise,try,assert},
keywordstyle={\color{Keywords}\bfseries},
morekeywords={[2]@invariant,pylab,numpy,np,scipy},
keywordstyle={[2]\color{Decorators}\slshape},
emph={self},
emphstyle={\color{self}\slshape},
}

\makeatletter
\@ifundefined{textcolor}{}
{%
 \definecolor{BLACK}{gray}{0}
 \definecolor{WHITE}{gray}{1}
 \definecolor{RED}{rgb}{1,0,0}
 \definecolor{GREEN}{rgb}{0,1,0}
 \definecolor{BLUE}{rgb}{0,0,1}
 \definecolor{CYAN}{cmyk}{1,0,0,0}
 \definecolor{MAGENTA}{cmyk}{0,1,0,0}
 \definecolor{YELLOW}{cmyk}{0,0,1,0}
}

\hypersetup{colorlinks=true,citecolor=blue,linkcolor=cyan,urlcolor=blue,filecolor= green, breaklinks=true}
\usepackage{url}
\usepackage{breakurl}

\makeatother

\begin{document}

\author{Lucas Friedrich\href{https://orcid.org/0000-0002-3488-8808}{\includegraphics[scale=0.05]{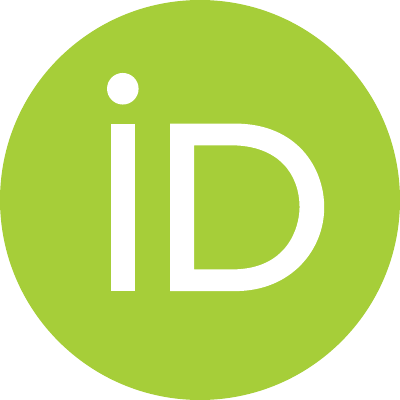}}}
\email{lucas.friedrich@acad.ufsm.br}
\affiliation{Physics Department, Center for Natural and Exact Sciences, Federal University of Santa Maria, Roraima Avenue 1000, Santa Maria, RS, 97105-900, Brazil}

\author{Jonas Maziero\href{https://orcid.org/0000-0002-2872-986X}{\includegraphics[scale=0.05]{orcidid.pdf}}}
\email{jonas.maziero@ufsm.br}
\affiliation{Physics Department, Center for Natural and Exact Sciences, Federal University of Santa Maria, Roraima Avenue 1000, Santa Maria, RS, 97105-900, Brazil}

\title{
Quantum neural network with ensemble learning \\ to mitigate barren plateaus and cost function concentration
}

\begin{abstract}
The rapid development of quantum computers promises transformative impacts across diverse fields of science and technology. Quantum neural networks (QNNs), as a forefront application, hold substantial potential. Despite the multitude of proposed models in the literature, persistent challenges, notably the vanishing gradient (VG) and cost function concentration (CFC) problems, impede their widespread success. In this study, we introduce a novel approach to quantum neural network construction, specifically addressing the issues of VG and CFC.
Our methodology employs ensemble learning, advocating for the simultaneous deployment of multiple quantum circuits with a depth equal to \(1\), a departure from the conventional use of a single quantum circuit with depth \(L\). We assess the efficacy of our proposed model through a comparative analysis with a conventionally constructed QNN. The evaluation unfolds in the context of a classification problem, yielding valuable insights into the potential advantages of our innovative approach.
\end{abstract}


\maketitle
\section{Introduction}

Machine learning is a subfield of artificial intelligence in which a computational model is trained using a dataset to identify underlying patterns in the data. In recent years, various fields of science and technology have benefited from these approaches such as computer vision \cite{computer_vision_1,computer_vision_2,computer_vision_3}, portfolio optimization \cite{ZHANG_portfolio_optimization}, chemical analysis \cite{Deep_Chemistry}, natural language processing \cite{nlp_1,nlp_2}, and drug development \cite{drog_discovery}. However, several challenges still persist, in this context, quantum machine learning emerges as a possible solution to overcome such difficulties.

Quantum machine learning is an interdisciplinary domain that combines the principles of quantum computing with machine learning. This emerging field aims to create machine learning models that, by exploiting quantum properties such as entanglement and superposition, seek to surpass classical models. One example is the ability to use the symmetry of training data to develop models that require fewer data \cite{geometric_QNN}, but still exhibit good generalization capabilities \cite{Generalization_few_data}.

Recently, a wide range of models has been proposed, including quantum neural networks \cite{quantum_model_multilayer_perception}, quantum kernel models \cite{kernel_methods}, quantum convolutional neural networks \cite{quantum_Convolutional}, and hybrid quantum-classical neural networks (HQCNN) \cite{hybrid_1, hybrid_2, hybrid_3, hybrid_4}. However, in general, all these models are based on variational quantum algorithms (VQAs) \cite{Cerezo_VQA}, which, in the era of noisy intermediate-scale quantum (NISQ) devices, emerge as the primary strategy for exploring potential quantum advantages.

VQAs consist of an interactive method in which a classical optimizer is employed to adjust the parameters $\pmb{\theta}$ of a quantum circuit in order to minimize a cost function $C$. The quantum circuit is initially constructed using a parametrization $V$, obtained via a set of quantum gates, with the purpose of preparing an initial state. Next, a parametrization $U$ is applied, also derived from a set of quantum gates, some of which depend on the parameters $\pmb{\theta}$ to be optimized. Finally, measurements are made to extract classical information from the system, allowing us to compute the cost function $C$.

Although VQAs are widely used in the construction of different quantum machine learning models, they face several challenges. One such challenge is the vanishing gradient problem, also known as Barren Plateaus (BPs) \cite{barrenPlateaus_0,barrenPlateaus_1,barrenPlateaus_2,barrenPlateaus_3,barrenPlateaus_4,barrenPlateaus_5,barrenPlateaus_6,barrenPlateaus_7,barrenPlateaus_8,barrenPlateaus_9,barrenPlateaus_10,barrenPlateaus_11,kashif_BP}. This problem arises from the tendency of the derivatives of the cost function with respect to any parameter $\theta_k$ to approach zero as the number of qubits and the depth of the parametrization increase, thus hindering the training of these models. Moreover, a second challenge is related to the concentration of the cost function (CFC) as the expressivity of the quantum circuit increases \cite{FRIEDRICH_expressiviy}, leading to the concentration of the cost function around a fixed value, which negatively impacts the trainability and applicability of VQAs.

In this article, we aim to introduce a new approach to constructing a quantum machine learning model, specifically an HQCNN model, which combines classical and quantum layers. More specifically, we propose that by rewriting the quantum layers not as a single quantum circuit with a depth-$L$ parameterization but rather as a set of $L$ quantum circuits with a depth-1 parameterization, we can mitigate the issues of BPs and CFC.

To demonstrate this, the article is organized as follows: in Section \ref{sec:qnn}, we begin with an introduction to VQAs, and in Subsections \ref{subsec:bp} and \ref{subsec:const_function}, we present the definitions of BPs and CFC, respectively. Next, in Section \ref{sec:hqcnn}, we provide a brief description of HQCNNs, followed by Subsection \ref{sec:Method}, where we describe the new HQCNN construction approach proposed in this article. In Section \ref{sec:Result}, we present some of the results obtained in this study. To this end, in Subsection \ref{subsec:ProblemDescription}, we describe the problem analyzed in this work, followed by the presentation of the obtained results in Subsection \ref{subsec:NumericalResults}, and we conclude with a discussion of these results in Subsection \ref{sec:Discussion}. Finally, in Section \ref{sec:Conclusion}, we present our final remarks.

\section{Variational quantum algorithms}
\label{sec:qnn}

In recent years, numerous studies have been conducted with the aim of developing quantum machine learning models \cite{quantum_model_multilayer_perception,kernel_methods,quantum_Convolutional,hybrid_1, hybrid_2, hybrid_3,hybrid_4}, which, by leveraging quantum properties such as entanglement and superposition, can theoretically surpass their classical counterparts. Despite various proposals, such as quantum neural networks \cite{quantum_model_multilayer_perception} and hybrid quantum-classical neural networks \cite{hybrid_1, hybrid_2, hybrid_3, hybrid_4}, all these models share a fundamental element: variational quantum algorithms.

Variational quantum algorithms represent an iterative approach in which a classical optimizer is used to adjust the parameters $\pmb{\theta}$ of a quantum circuit, aiming to minimize a cost function $C$. The objective is to determine $\pmb{\theta}_{\text{opt}}$ such that:
\begin{equation} 
\pmb{\theta}_{\text{opt}} = \arg \min_{\pmb{\theta}} C(\pmb{\theta}). 
\end{equation}
In general, a quantum circuit can be divided into three distinct parts. The first is the parameterization $V$, composed of a set of quantum logic gates responsible for preparing the initial state. In the context of quantum machine learning, this initial state corresponds to encoding the input data into a quantum state. Thus, given a training dataset $\mathcal{D} = \{\pmb{x}_{i}, y_{i}\}_{i=1}^{N}$, where $\pmb{x}_{i}$ represents the input data and $y_{i}$ their respective labels, the parameterization $V$ is used to generate:
\begin{equation} 
|\pmb{x}_{i}\rangle = V(\pmb{x}_{i})|0\rangle^{\otimes n}, 
\end{equation}
where $n$ is the number of qubits used by the quantum circuit. As evidenced in several studies \cite{data_encoding_SCHULD}, the choice of $V$ is crucial for the performance of quantum machine learning models, and despite the existence of different encoding methods \cite{robust_data_encoder,Data_re_uploading}, this remains an open area of research.

The second part of the quantum circuit consists of the parameterization $U$, which is also composed of a set of quantum logic gates, some of which depend on the parameters $\pmb{\theta}$ to be optimized. In general, $U$ is defined as:
\begin{equation} 
U(\pmb{\theta}) = \prod_{l=1}^{L} U_{l}(\pmb{\theta}_{l}), \label{eq:parametrization}
\end{equation}
where $U_{l}$ represents an arbitrary unitary operation, and $L$ indicates the depth of this parameterization. These unitary operations are often implemented with rotation and CNOT gates. Although the set of gates used may vary for each unitary operation, in general, they all use the same set, differing only in the parameters $\pmb{\theta}_{l}$ employed by each set.

The third and final part of the quantum circuit comprises the measurements performed at the end of the process. These measurements are typically defined as the expectation value of an observable $O$, that is:\begin{equation} 
f = \langle O \rangle = \text{Tr}\big[O  U(\pmb{\theta}) |\pmb{x}_{i}\rangle \langle \pmb{x}_{i}| U(\pmb{\theta})^{\dagger}\big]. \label{eq:media}
\end{equation}
The purpose of these measurements is to extract classical information from the quantum circuit. The interpretation of this information depends on the context in question. For example, in HQCNN models, such information may be used as input for other layers, whether classical or quantum. Alternatively, this information can be treated as the cost function itself, i.e., $f = C$.

The training of these models occurs iteratively, with the optimization of parameters $\pmb{\theta}$ being carried out to minimize the cost function. Although several optimization methods have been proposed \cite{xie_grad}, gradient descent is widely used. In this method, the parameters of the quantum circuit are adjusted based on the gradient of the cost function, according to the update rule:\begin{equation} 
\pmb{\theta}_{t+1} = \pmb{\theta}_{t} - \eta \nabla_{\pmb{\theta}_{t}} C, 
\end{equation}
where $t$ denotes the current iteration (epoch) and $\eta$ is the learning rate. In the case of quantum circuits, the gradient is calculated using the method known as the parameter shift rule \cite{Schuld_grad, Crooks_grad}, whose formula is given by:\begin{equation} 
\partial_{k} f = \frac{1}{2}\bigg[f\bigg(\theta_{k} + \frac{\pi}{2}\bigg) - f\bigg(\theta_{k} - \frac{\pi}{2}\bigg)\bigg]. 
\end{equation}

As one can observe, the computational cost associated with the parameter shift rule grows linearly with the number of parameters to be optimized. Although this cost is negligible in models with few parameters, it becomes more significant as the number of parameters increases, potentially making optimization unfeasible in extreme cases. In addition to the high computational cost, these models face other significant challenges, such as gradient disappearance and cost function concentration, which will be discussed below.

\subsection{Barren plateaus}\label{subsec:bp}

Currently, one of the main challenges faced by VQAs and various quantum machine learning models is the BPs problem. This problem was initially reported in Ref. \cite{barrenPlateaus_0}, where it was found that the average value of the derivative of a function, defined similarly to the one described in Eq.\eqref{eq:media}, becomes zero for any $\theta_k$, while its variance tends to decrease exponentially as the number of qubits in the quantum circuit increases. In other words, the BPs problem can be defined as follows:

\begin{definition}
 
Consider the function defined in Eq. \eqref{eq:media} with \(O\) being any observable and the parameterization \(U\) given by Eq. \eqref{eq:parametrization}. This function exhibits barren plateaus if the variance of the partial derivative of the function \(f\) with respect to any parameter \(\theta_{k}\) vanishes exponentially with the number of qubits \(n\). That is,
\begin{equation}
    Var[\partial_{k}f] \leqslant  G(n), \text{ with } G(n) \in \mathcal{O}\bigg(\frac{1}{b^{n}}\bigg),
\end{equation}
where \(b>1\).
\label{definition:2}
\end{definition}

This definition is based on Chebyshev's inequality, which states that
\begin{equation}
    Pr( |\partial_{k}f -  \langle \partial_{k}f \rangle| \geqslant \delta )  \leqslant \frac{ Var(\langle \partial_{k}f \rangle)  }{\delta^{2}}, \label{eq:Chebyshev}
\end{equation}
i.e., the probability that $\partial_{k}f$ deviates from its mean $\langle \partial_{k}f \rangle$ by a value greater than or equal to $\delta$ is upper-bounded by the variance of $\partial_{k}f$. However, since the mean value of $\partial_{k}f$ is zero and its variance decreases exponentially as the number of qubits in the quantum circuit increases, it follows that the probability of $\partial_{k}f$ deviating from zero also tends to decrease exponentially. In other words, for a quantum circuit with a sufficiently large number of qubits, the derivatives vanish. Consequently, the efficient optimization of the model's parameters becomes unfeasible, as this optimization process relies directly on these derivatives.

One of the factors contributing to the issue of BPs is the selection of the cost function. As demonstrated in Ref. \cite{barrenPlateaus_1}, there are two ways to define our cost function: the global cost function and the local cost function. The global cost function is obtained when all qubits are measured simultaneously; in this scenario, our model is consistently impacted by BPs. Conversely, the local cost function is implemented when qubits are measured individually or in pairs. In this case, there are instances where our model is not susceptible to the problem of BPs, particularly when the relationship between the depth of the parameterization and the number of qubits is \(\mathcal{O}(1)\) or \(\mathcal{O}(\log(n))\). Additionally, BPs have been linked to various other factors \cite{barrenPlateaus_2, barrenPlateaus_3, barrenPlateaus_4, barrenPlateaus_5, barrenPlateaus_6}. Consequently, several methods have been proposed to alleviate BPs \cite{barrenPlateaus_7, barrenPlateaus_8, barrenPlateaus_9, barrenPlateaus_10, barrenPlateaus_11,kashif_BP}, but this remains an ongoing area of research.

\subsection{Concentration of the cost function}
\label{subsec:const_function}

Another challenge faced by VQAs and, consequently, by quantum machine learning models, is the problem of cost function concentration. This problem arises from the tendency of the function, defined in Eq. \eqref{eq:media}, to converge to a fixed value. Interestingly, the problem of BPs can be seen as a manifestation of this concentration of the cost function. However, as presented in Ref.  \cite{FRIEDRICH_expressiviy}, cost function concentration is deeply related to the expressiveness of the parametrization. This result is particularly relevant because, even if we manage to develop an optimization method that is not affected by the BP problem, VQAs would still be impacted by the cost function concentration issue, which would substantially hinder their training capability and applicability.

\begin{figure}[ht]
    \centering
    \includegraphics[scale=0.45]{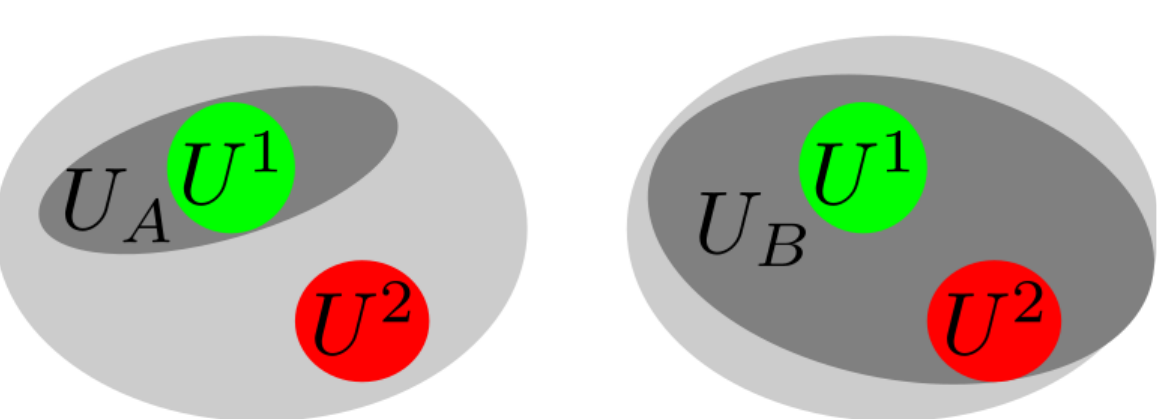}
    \caption{
    In this figure, we illustrate how expressibility can be interpreted as the number of unitaries $U^{i}$ accessed by a given parameterization $U_{S}$. We represent the solutions to two problems, $A$ and $B$, by the unitaries $U^{1}$ and $U^{2}$, respectively. Our goal is to obtain a parameterization $U_{S}$ that can access these two unitaries. The accessible space is depicted in dark gray for parameterizations $U_{A}$ and $U_{B}$. As we can observe, while $U_{A}$ can only access $U^{1}$, $U_{B}$ can access both solutions. In this case, we say that the expressibility of $U_{B}$ is greater than that of $U_{A}$.}
    \label{fig:expr}
\end{figure}

To grasp the concept of expressibility, consider the scenario depicted in Fig. \ref{fig:expr}. We aim to devise a parameterization \(U\) capable of solving both problem \(A\) and problem \(B\). The solution to problem \(A\) is encapsulated by a unitary denoted as \(U^{1}\), while problem \(B\) is solved by another unitary termed \(U^{2}\). Our objective is to devise a parameterization \(U_{S}\) that can achieve both \(U^{1}\) and \(U^{2}\). However, as illustrated, we can formulate this parameterization using distinct sets of quantum gates. Suppose we develop a parameterization \(U_{A}\) utilizing a particular set of gates and another parameterization \(U_{B}\) employing a different set of gates. 
Now, consider that while employing the parameterization \(U_{A}\), we can solely access the solution \(U^{1}\) for problem \(A\) but not the solution for problem \(B\). Conversely, when utilizing the parameterization \(U_{B}\), we can access the solution for both problems. In this scenario, we denote that the expressiveness of parameterization \(U_{B}\) surpasses that of parameterization \(U_{A}\). Consequently, expressiveness can be conceptualized as the breadth of functions—or in this context, the array of solutions—that a specific parameterization can access.

Mathematically, expressibility is defined as:
\begin{equation}
    A_{\mathbb{U}}^{t}(.)  := \int_{\mathcal{U}(d)}d\mu (V)V^{\otimes t}(.)(V^{\dagger})^{\otimes t} - \int_{\mathbb{U}}dUU^{\otimes t}(.)(U^{\dagger})^{\otimes t},
     \label{eq:expr_definition}
\end{equation}
where \(d\mu(V)\) is a volume element of the Haar measure, and \(dU\) is a volume element corresponding to the uniform distribution over \(\mathbb{U}\). Given this definition and the function defined in Eq.~\eqref{eq:media}, it was shown in Ref. \cite{FRIEDRICH_expressiviy} that:
\begin{equation}
        \bigg| E_{\mathbb{U}}[f] - \frac{Tr[O]}{d} \bigg| \leqslant \| O \|_{2} \| A(\rho) \|_{2}.\label{eq:eq_teo_1}
\end{equation} 
This result indicates that the average value\footnote{In this work, we adopt two distinct notations to represent the mean value. That is, the mean value of a function $f$ with respect to $U$ can be expressed equivalently as \(E_{\mathbb{U}}[f]\) or $\langle f \rangle = \langle f \rangle_{U}$.
}  \(E_{\mathbb{U}}[f]\)
of the function defined in Eq. \eqref{eq:media} will concentrate around \(Tr[O]/d\) as the expressiveness increases, because \(\| A(\rho) \|_{2} \rightarrow 0\) as the parameterization becomes more expressive.

This has serious implications. Suppose we need to solve a classification problem. Initially, one might expect that the best model would be the one with maximum expressiveness, as it would then be able to access all possible solutions. However, this result implies that in this case, the cost function defined in Eq.~\eqref{eq:media} would concentrate around \(Tr[O]/d\). Therefore, let us assume that \(Tr[O]/d = 0\) and that we are working with a classification problem where the labels are \( (0,1) \). In this scenario, it would be impossible to obtain the label \( 1 \). Consequently, our model would be incapable of solving the classification task.

\section{Hybrid quantum-classical neural networks}
\label{sec:hqcnn}

\begin{figure}[ht]
    \centering
    \includegraphics[scale=0.8]{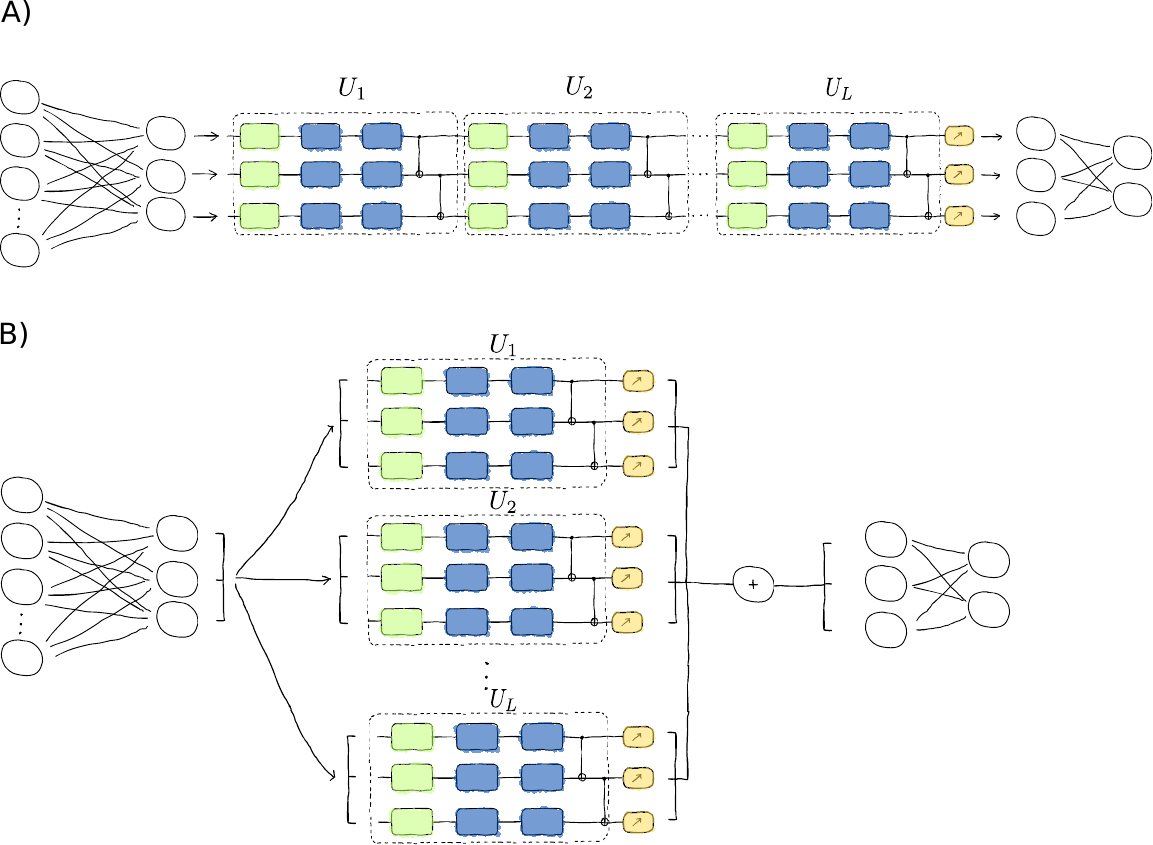}
    \caption{ 
    Illustration of a hybrid quantum-classical neural network (HQCNN). A) This illustration shows an HQCNN model where two classical layers are followed by a quantum layer, and finally, another two classical layers are applied. In this example, the quantum layer is obtained using a single quantum circuit with depth $L$. The gates used to encode the data obtained from the classical layer are highlighted in green. The gates depending on parameters to be optimized are shown in blue. B) Illustration of an HQCNN model using the new method. In this example, the model consists of two classical layers and one quantum layer. In the quantum layer, instead of using a single quantum circuit of depth $L$, we employ $L$ quantum circuits of depth $1$.}
    \label{fig:modeloHibrido}
\end{figure}

In this study, our analysis will focus on the HQCNN models, as illustrated in Fig. \ref{fig:modeloHibrido}. These models combine classical and quantum layers, leveraging the advantages inherent to both approaches. In the NISQ era, quantum computers face substantial limitations, such as the restricted number of qubits and the operational constraints imposed by current technology. In this context, the HQCNN models emerge as a promising solution, as they allow for the flexible configuration of the number of classical and quantum layers, as well as the control of data flow between them. This flexibility enables the adaptation of the models to existing technological limitations.

In Fig. \ref{fig:modeloHibrido} A, an example of a typical HQCNN model is presented. This model consists of two classical layers\footnote{The classical layers used in this model are described by the equation $f(\pmb{x}) = \phi(\pmb{x}\pmb{w} + \pmb{b})$, where $\{\pmb{w}, \pmb{b}\}$ represent the parameters that need to be optimized, and $\phi(\cdot)$ denotes the activation function. In the case of the first layer, the ReLU function was used, while the softmax function was chosen for the last.}, placed at the beginning and end of the model, and one quantum layer, placed between the classical layers. The first classical layer processes the input data so that the dimension of its output matches the number of qubits used in the quantum layer, while the second classical layer processes the data from the quantum layer and returns an output. The quantum layer, on the other hand, consists of a quantum circuit whose mathematical description is given by:
\begin{equation}
    y_{j} = \langle O_{j} \rangle = Tr[O_{j} U | \pmb{0} \rangle \langle \pmb{0} |U ],\label{eq:hqcnn_1}
\end{equation}
where $O_{j} := \mathbb{I}_{\bar{j}} \otimes |0\rangle \langle 0|$, with the index $\bar{j}$ indicating that the identity matrix will be applied to all qubits except the one with index $j$, where $|0\rangle \langle 0|$ will be applied. Finally, $U$ is the parametrization used, and its definition is given by:
\begin{equation} 
U = \prod_{l=1}^{L}U_{l}(\pmb{\theta}_{l})V(\pmb{x}_{i}) := \prod_{l=1}^{L}U_{l},\label{eq:hqcnn_2}
\end{equation}  
where $V$, defined as
\begin{equation}
    V = \bigotimes_{k=0}^{n-1}R_{y}(x_{i}^{k}),
\end{equation}
is a parametrization used to encode the input data $\pmb{x}_{i} \in \mathbb{R}^n$, which in this case are the outputs obtained from the first classical layer, and $U_{l}(\pmb{\theta}_{l})$ is an arbitrary parametrization that depends on the parameters $\pmb{\theta}_{l}$ to be optimized. This encoding method, where we use $V$ intercalated with the parametrizations $U_{l}(\pmb{\theta}_{l})$, is known as data re-uploading and was initially proposed in Ref. \cite{Data_re_uploading}. Furthermore, the parameterization $V$, used to encode the input data, is obtained from the tensor product of the $n$ rotation gates $R_y$. In order to avoid issues related to multiples of $2\pi$, that is, distinct values that, being multiples of $2\pi$, are treated as identical, we normalize the input data to the interval $[0,2\pi]$ before applying it to the parametrization $V$.

\subsection{Method}\label{sec:Method}

In this subsection, we present the proposed new method. As illustrated in Fig. \ref{fig:modeloHibrido} A, a typical HQCNN model combines both classical and quantum layers, where, in this specific case, the model consists of two classical layers and one quantum layer. The quantum layer is obtained through a single quantum circuit, as described by Eq. \eqref{eq:hqcnn_1}, with $U$ defined by Eq. \eqref{eq:hqcnn_2}. As can be seen in Eq. \eqref{eq:hqcnn_2}, $U$ is generated from the product of $L$ layers $U_{l} = U_{l}(\pmb{\theta}_{l})V(\pmb{x}_{i})$. As highlighted in several studies \cite{barrenPlateaus_1,FRIEDRICH_expressiviy}, in general, the deeper the depth $L$, the greater the impact of the BP and CFC problems. Therefore, to mitigate the issues related to BP and CFC, we propose the use of the technique known as \textit{ensemble learning} to reconstruct the quantum layer.

\textit{Ensemble learning} is a machine learning approach that combines multiple models to improve the accuracy and robustness of predictions compared to individual models. However, unlike the conventional approach, in which we combine several models, we propose to apply this technique exclusively to the quantum layer. More specifically, instead of obtaining the quantum layer using only a single quantum circuit described by Eq. \eqref{eq:hqcnn_1} with $U$ defined by Eq. \eqref{eq:hqcnn_2}, the quantum layer will now be obtained by combining $L$ quantum circuits, each described by Eq. \eqref{eq:hqcnn_1}, but with $U$ generated from a single layer $U_{l} = U_{l}(\pmb{\theta}_{l})V(\pmb{x}_{i})$, that is, $L = 1$. Fig. \ref{fig:modeloHibrido} B illustrates this newly proposed method. Finally, from the output of each quantum circuit, which will be represented by a vector $\pmb{y}_l$, we compute 
\begin{equation}
\pmb{Y} = \sum_{l=1}^{L}\pmb{y}_{l},\label{eq:sumYl}
\end{equation}
and use this result as input for the subsequent layer.

Additionally, a potential benefit of this new method is the mitigation of noise effects, a central challenge in the NISQ era. Current quantum computers are affected by noise arising from undesirable interactions with the environment, imperfections in the qubits, and technological limitations. The level of noise in a quantum circuit is closely related to the number of logic gates used: the greater the number of gates, the more pronounced the cumulative effect of noise, compromising the accuracy of the results. Thus, by reducing the depth of the parametrization – and consequently the number of logic gates – we can minimize the cumulative effects of noise and improve the accuracy of the results generated by the model.

\section{Result}
\label{sec:Result}

In this section, we present the results obtained in this study. To achieve them, we used the PyTorch \cite{Paszke} and PennyLane \cite{Pennylane} libraries for the construction of both the standard HQCNN model and the newly proposed model. PyTorch is widely recognized as one of the leading libraries for the development of machine learning models. In turn, although there are currently some libraries that facilitate the development of quantum machine learning models \cite{quforge,sQUlearn}, PennyLane is 
the most used by the community.

Initially, in Sec. \ref{subsec:ProblemDescription}, a description will be provided for both the models used in this study and the problem that will be applied to validate the new method. Then, in Sec. \ref{subsec:NumericalResults}, some of the results obtained will be presented.

\subsection{Problem Description}
\label{subsec:ProblemDescription}

As can be seen in Eq. \eqref{eq:hqcnn_2}, the quantum circuit depends on the choice of a parameterization $U_{l}(\pmb{\theta}_{l})$. This parameterization is defined through a sequence of quantum logic gates, where all or some of them depend on the parameters to be optimized. Although the choice of gates used can encompass all possible gates, they are usually restricted to rotation and CNOT gates.

\begin{figure}[ht]
    \centering
    \includegraphics[width=1\linewidth]{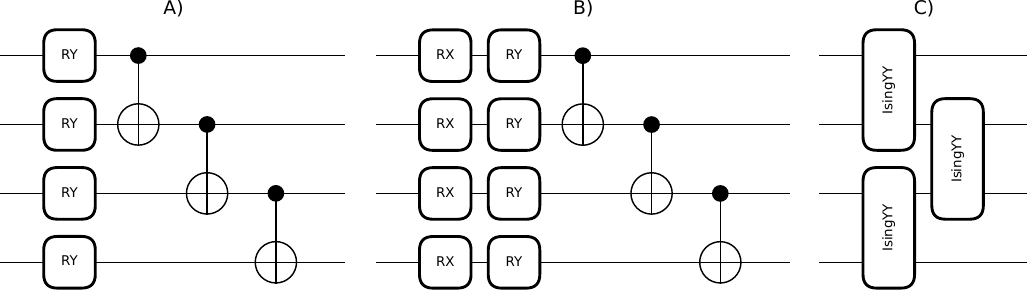}
    \caption{Illustration of the three parametrizations $U_l(\pmb{\theta}_l)$ used in this study. In this context, IsingYY refers to the two-qubit rotation gate, defined as $R_{YY}(\theta) = e^{-i \frac{\theta}{2} Y \otimes Y}$.
    }
    \label{fig:models}
\end{figure}

Even considering only rotation and CNOT gates, there is a large number of possible combinations that can be used. Given this, in this study, three possible configurations for $U_{l}(\pmb{\theta}_{l})$ will be considered, which are presented in Fig. \ref{fig:models}. This will allow us to conduct our analyses in a more comprehensive manner compared to an approach that considers only a single configuration for $U_{l}(\pmb{\theta}_{l})$.

To validate the new method, the behavior of the variance of the partial derivative with respect to an arbitrary parameter will initially be analyzed. For this purpose, a comparison will be made between the variance obtained using the standard approach, in which the quantum layer is generated by a single quantum circuit, and the new method, in which $L$ quantum circuits with a depth of 1 are used. The variance will be computed based on a sample of 1000 values for each model. In other words, for each model, the derivative calculation will be performed 1000 times using different parameters $\pmb{\theta}$.

Furthermore, to obtain the variance, it was considered that $L = n$, meaning that the number of layers, as defined in Eq. \eqref{eq:hqcnn_2}, grows linearly with the number of qubits. According to Ref. \cite{barrenPlateaus_1}, this condition implies the presence of the barren plateaus (BPs) problem.

Furthermore, in the case of the standard model, the derivative will be calculated with respect to the $n$ expectation values $\langle O_{j} \rangle$, while in the case of the new method, the derivative will be computed with respect to the mean of $\pmb{Y}$, described by Eq. \eqref{eq:sumYl}. It is important to emphasize that this approach is viable only because the expectation values $\langle O_{j} \rangle$ of the standard model and the components of the vector $\pmb{Y}$, obtained from the sum of the $L$ expectation values of each quantum circuit in the new model, are linearly independent.

Besides the variance, to verify the validity of the new method, the performance of both the standard HQCNN model and the model using ensemble learning will be analyzed in a multi-class classification problem. More specifically, in this study, the MNIST dataset will be used, one of the most well-known and widely used collections in research on image classification models. MNIST consists of 70,000 images of handwritten digits, divided into two subsets: 60,000 images for training and 10,000 for testing. Each image represents a handwritten digit ranging from 0 to 9 and is stored in a $28 \times 28$ pixel matrix, resulting in a vector of 784 features (or attributes). The images are grayscale, with pixel values ranging from 0 (black) to 255 (white). To facilitate processing, it is common to normalize these values so that the pixels range from 0 to 1, a procedure adopted in this study.

In traditional machine learning models, it is common to use the complete dataset, that is, all 60,000 training images and 10,000 test images, providing a robust foundation for model development and validation. However, in the context of quantum machine learning, where computational resources are limited and simulation on classical computers is required, the computational cost of training and testing models with the full MNIST dataset can be significant.

For this reason, it is common to use reduced versions of the dataset, with a smaller number of images. Thus, in this work, the classification task was simplified by using only a fraction of the MNIST. The dataset was restricted to digits 0, 1, and 2, thereby reducing the computational cost without losing the relevance of the classification problem. More concretely, the training set used in this study consists of 4,000 images, while the test set contains 400 images.

Furthermore, for the training of both models, we used the Adam optimizer with a learning rate set to $\eta = 0.001$. This specific value was selected based on its optimized performance with respect to the cost function and accuracy during preliminary simulations. Although this choice contributes to greater interpretability of the results, it is important to note that other values could be considered in order to explore possibilities for further optimization. Given the primary goal of introducing this new method in the construction of quantum machine learning models and performing a comparative analysis with the conventional approach, we deemed this value appropriate, without the need for exhaustive exploration. Additionally, we used the mean squared error (MSELoss) function, provided by PyTorch, as the cost function. Finally, in order to analyze the influence of parameter initialization on the obtained results, we repeated the simulations ten times for each model and hyperparameter set.
Below, some of the results obtained in this study are presented.

\subsection{Numerical Results}\label{subsec:NumericalResults}

\begin{figure}[H]
    \centering
    \includegraphics[width=1\linewidth]{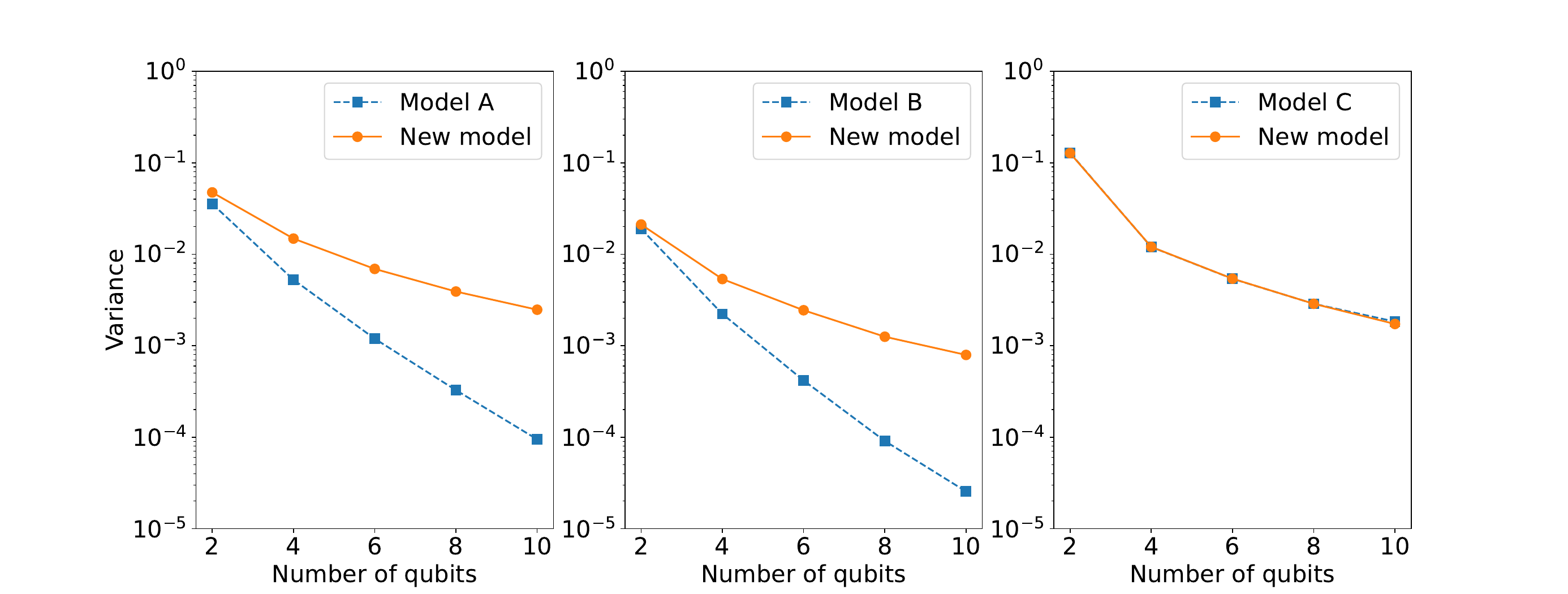}
    \caption{Variance analysis of the partial derivative with respect to the quantum layer used by the standard HQCNN model and the HQCNN model based on the new method. In the left plot, the behavior of the variances is observed when using the parametrization presented in Fig. \ref{fig:models} A. In the central plot, the variances corresponding to the parametrization in Fig. \ref{fig:models} B are shown. Finally, in the right plot, the behavior of the variances obtained when employing the parametrization in Fig. \ref{fig:models} C is illustrated. To obtain these results, the depth $L$ was scaled linearly with the number of qubits, that is, $L = n$.
    }
    \label{fig:model_var}
\end{figure}

Initially, Fig. \ref{fig:model_var} presents a comparison of the variance behavior when using the standard HQCNN model and the modified version based on the new approach for the three parameterizations illustrated in Fig. \ref{fig:models}. In Fig. \ref{fig:model_var}, each standard HQCNN model is referred to as Model A, Model B, or Model C, where the letter indicates the adopted parameterization. The modified version of HQCNN, built with the new approach, is simply called New Model, maintaining the correspondence with the utilized parameterization.

The results presented in Fig. \ref{fig:model_var} indicate that when using parameterizations A and B, the new approach indeed provides an increase in variance compared to the standard approach, thus allowing the mitigation of the BP problem. However, surprisingly, when parameterization C is adopted, the variance observed for both the standard HQCNN model and the version with the new approach are similar.

Although parameterization C does not allow the mitigation of the BP problem, as observed in Ref. \cite{FRIEDRICH_expressiviy}, expressivity is closely related to the depth of the parameterization. In general, the greater the depth, the higher the expressivity of the model. Therefore, despite not being able to mitigate the BP problem in this specific case, it is still possible to mitigate the problem of cost function concentration.

\begin{figure}[H]
    \centering
    \includegraphics[width=1\linewidth]{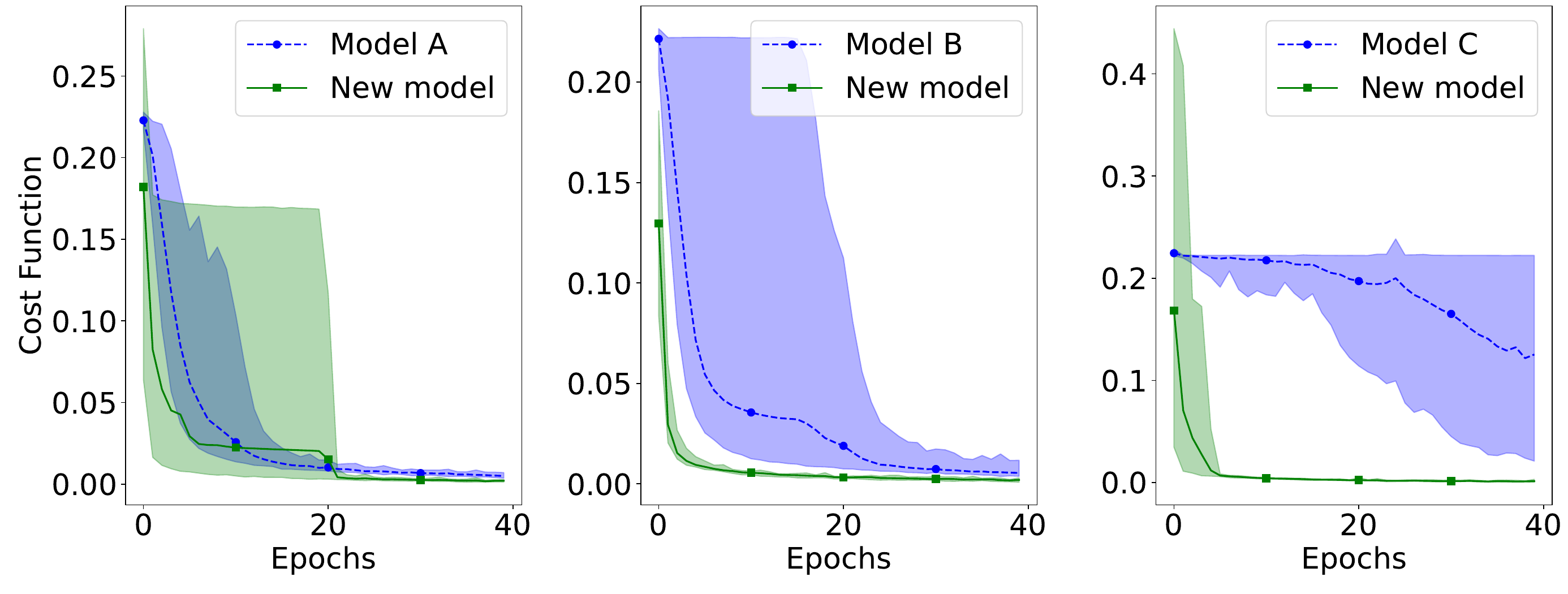}
    \caption{Comparative analysis of the cost function behavior when using the standard method and the new method. To obtain these results, quantum circuits with 8 qubits and depth $L = 40$ were employed in the case of the standard method, while for the new method, 40 quantum circuits were used, each with 8 qubits and depth $L = 1$. It is observed that, in all cases, the performance of the cost function when using the new method is superior to that obtained with the standard method.
    }
    \label{fig:loss_n_8_l_40}
\end{figure}

\begin{figure}[H]
    \centering
    \includegraphics[width=1\linewidth]{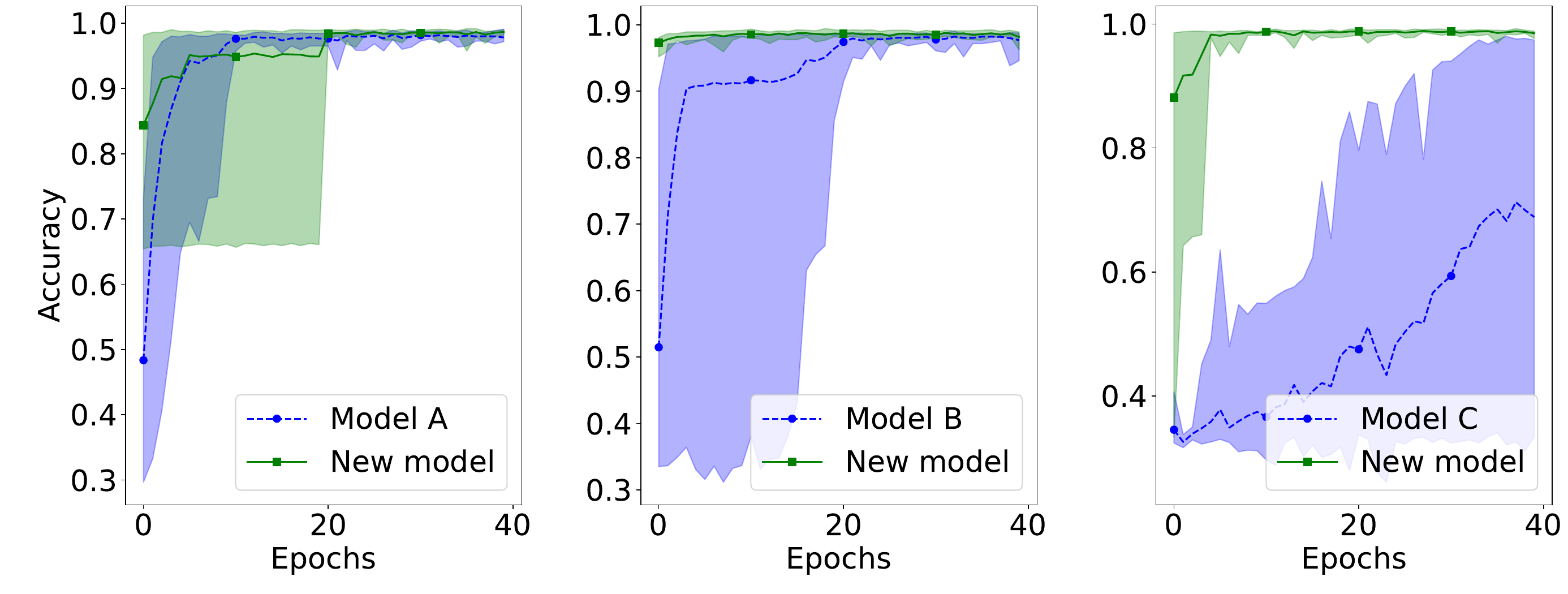}
    \caption{Accuracy analysis with respect to the test data. It is observed that, for parametrization B (central plot), the accuracy obtained by the HQCNN model using the new method quickly converges to a value close to the ideal.}
    \label{fig:acc_n_8_l_40}
\end{figure}

The Figs. \ref{fig:loss_n_8_l_40} and \ref{fig:acc_n_8_l_40} show, respectively, the behavior of the cost function and the accuracy for the classification problem considered in this study. From these figures, the following observations can be made:
\begin{enumerate}
    \item The choice of parameterization $U_{l}(\pmb{\theta}_l)$ directly influences the behavior of the cost function.
    
    \item The behavior of the cost function, for both the standard HQCNN model and the HQCNN model with the new approach, is impacted by parameter initialization.
    
    \item In general, the results obtained with the new approach are superior to those obtained with the standard model.
    
    \item For parameterization C, the results obtained with the new approach surpass those of the standard model, even if the BP problem is not mitigated.
    
    \item For parameterizations B and C, the HQCNN model using the new approach converges faster than the standard model.
    
    \item The accuracy obtained with parameterization B quickly converges to a value close to the optimal one.
\end{enumerate}

Additionally, in order to analyze how the number of qubits and the depth of the parametrization can influence the performance of the new method in comparison with the conventional method, we performed additional simulations in which we evaluated both the behavior of the cost function and the accuracy for models where the number of qubits and the depth of the parametrization vary. More specifically, we considered models with 4, 5, and 6 qubits, as well as with 2, 4, and 6 layers. The results obtained can be found in the Supplementary Information.

\subsection{Discussion}
\label{sec:Discussion}

The results obtained in this study demonstrate that the new method presents significant improvements over the standard method. First, as discussed in Ref. \cite{FRIEDRICH_expressiviy}, large values of expressivity, which leads to the concentration of the cost function, are closely related to the depth of the parametrization. Thus, since in the new method this depth is fixed at $L=1$, it becomes possible to mitigate this issue.

Regarding the problem of BPs, the variance analysis presented in Fig. \ref{fig:model_var} revealed that, for two of the evaluated parametrizations, specifically parametrization A and parametrization B, the adoption of the new method allowed the mitigation of this problem. However, for parametrization C, the BP problem was not mitigated. This result is surprising since, according to Ref. \cite{barrenPlateaus_1}, for quantum circuits that use local measurements, as long as the depth of the parametrization scales as $\mathcal{O}(1)$ or $\mathcal{O}(\log(n))$ with the number of qubits, the model should not suffer from the BP problem.

A possible explanation for this discrepancy concerning the result of Ref. \cite{barrenPlateaus_1} lies in the fact that such relations were obtained assuming that the parametrization used forms a $2$-design. However, in the present study, particularly for parametrization C, this criterion was not strictly followed. In other words, parametrization C does not satisfactorily approximate a $2$-design, which justifies the observation of contradictory results. In fact, in Ref. \cite{barrenPlateaus_1}, to demonstrate the relationship between the choice of measurement and the BP problem, the authors used models substantially larger than those employed in this study. Therefore, based on the results presented in Fig. \ref{fig:model_var}, we can conclude that the effectiveness of the new method in mitigating BPs will depend on how closely the chosen parametrization approximates a $2$-design.

Regarding the performance of the methods in the classification problem, the new method using parametrization A presented a slightly lower cost function than that of the HQCNN model built using the standard method. However, a greater dependence on parameter initialization was observed during the initial training epochs. This dependence, however, disappears after a certain number of iterations, indicating that despite the initial sensitivity, after a certain number of iterations, the initialization of the parameters ceases to be significant for the model's performance.

In the other two cases analyzed, involving parametrizations B and C, the HQCNN model based on the new method demonstrated significantly superior performance, achieving considerably lower cost function values compared to the standard method. Furthermore, dependence on parameter initialization was observed only in the standard method, whereas when employing the new method for constructing the HQCNN model, we observed a more stable and robust learning process.

Finally, when analyzing the behavior of both methods under variations in the number of qubits and the depth of the parametrization (see Supplementary Information), we found that although the new method contributes to mitigating the BP problem and the concentration of the cost function, its performance will depend on the appropriate choice of hyperparameters. These, in turn, will be crucial in determining the impact of parameter initialization on the obtained results. This aspect was already expected, given that, in machine learning problems, both classical and quantum, the selection of hyperparameters plays a fundamental role in model performance.

\section{Conclusion}
\label{sec:Conclusion}

In this study, our aim was to showcase how \textit{ensemble learning} can be harnessed to tackle the challenges of gradient vanishing and cost function concentration in quantum neural networks. Through a comparative analysis, we juxtaposed the performance of a traditionally constructed model with one crafted via \textit{ensemble learning} for a classification task. The results underscore that by leveraging this method we can develop models that not only generally match the performance of conventional ones but also effectively mitigate these challenges. This is attributable to the significant reduction in parameterization complexity facilitated by \textit{ensemble learning}. Consequently, we have presented an alternative avenue for constructing quantum neural network models capable of alleviating both issues.


\begin{acknowledgments}
This work was supported by the Coordination for the Improvement of Higher Education Personnel (CAPES) under Grant No. 88887.829212/2023-00, by the National Council for Scientific and Technological Development (CNPq) under Grants No. 309862/2021-3, No. 409673/2022-6, and No. 421792/2022-1, and by the National Institute for the Science and Technology of Quantum Information (INCT-IQ) under Grant No. 465469/2014-0.
\end{acknowledgments}

\vspace{0.1cm}

\textbf{Data availability.}
The numerical data and code generated in this work is available at \url{https://github.com/lucasfriedrich97/qnnEnsemble}.

\vspace{0.1cm}

\textbf{Contributions} The project was conceived by L.F., who also carried out the numerical simulations. J.M. supervised the research. L.F. wrote the first version of the article, which was revised by J.M.. 

\vspace{0.1cm}

\textbf{Competing interests.}
The authors declare no competing interests.


\widetext
\newpage

\begin{center}
\vskip0.5cm
{\Large \textbf{Supplementary information for ``Quantum neural network with ensemble learning to mitigate barren plateaus and cost function concentration''}}
\vskip0.2cm
Lucas Friedrich, and Jonas Maziero
\vskip0.1cm
\textit{Physics Departament, Center for Natural and Exact Sciences, \\
Federal University of Santa Maria, Roraima Avenue 1000, 97105-900, Santa Maria, RS, Brazil}
\vskip0.1cm
\end{center}

\setcounter{equation}{0}
\setcounter{figure}{0}
\setcounter{page}{1}
\renewcommand{\thefigure}{S\arabic{figure}}
\renewcommand{\theequation}{S\arabic{equation}}

In the supplementary information, we present additional simulation results for the classification problem considered in this study. Specifically, we performed a comparison between the HQCNN model using the standard method, in which the quantum layer consists of a single quantum circuit with a parametrization of depth $L$, and the HQCNN model based on the new method proposed in this work. In the latter case, we employed the technique known as \textit{ensemble learning} to construct the quantum layer, using $L$ quantum circuits, each with a parametrization of depth equal to 1, but now considering circuits with different numbers of qubits, denoted by NQ (specifically, 4, 5, and 6), and different depths, denoted by L (specifically, 2, 4, and 6).  

In Section~\ref{SI:CostFunction}, we presented the analysis of the behavior of the cost function with respect to the training data, while in Section~\ref{SI:Accuracy}, we discuss the accuracy behavior with respect to the test data.

\section{Cost function}\label{SI:CostFunction}

\begin{figure}[H]
    \centering
    \includegraphics[width=1\linewidth]{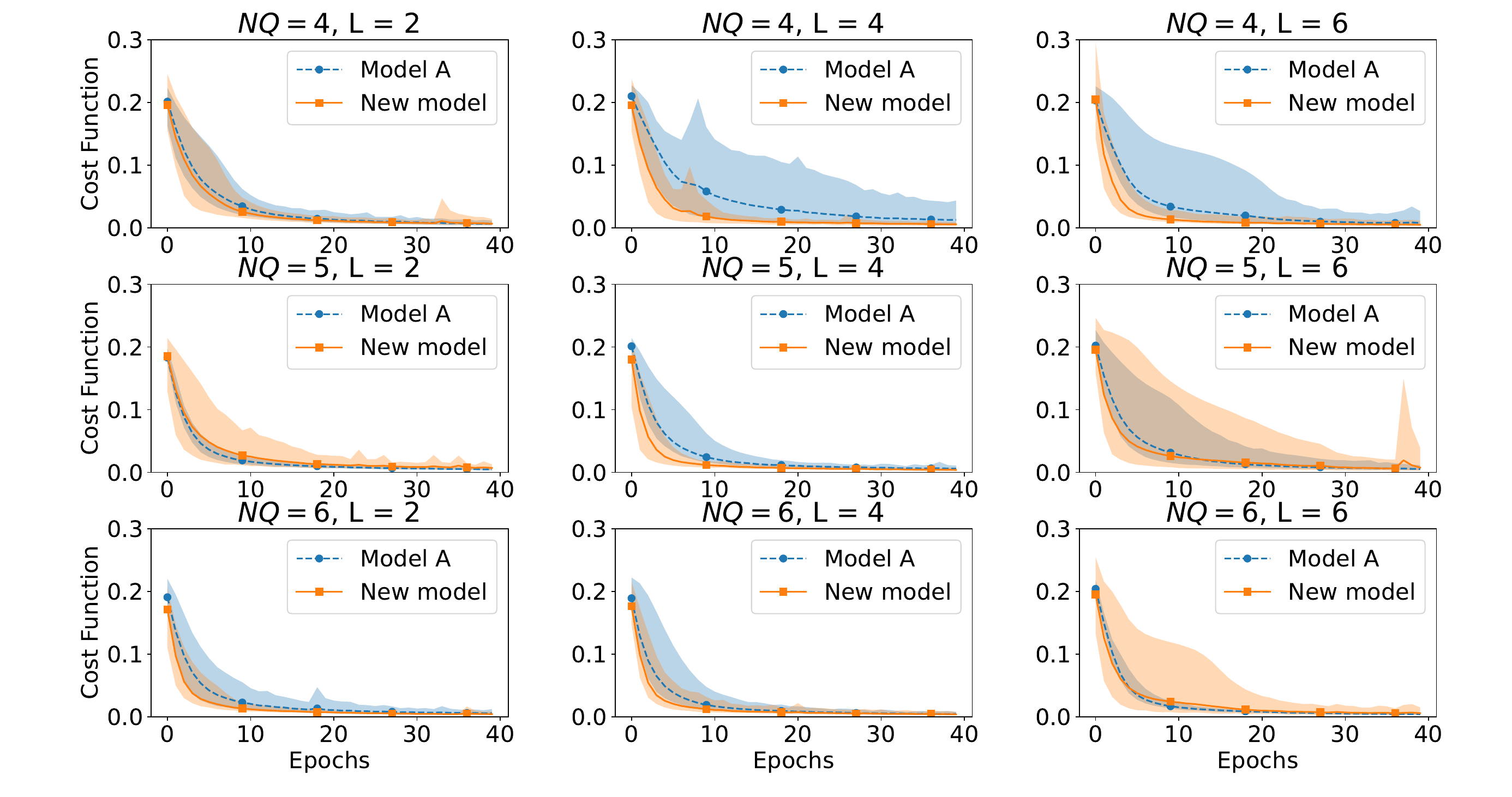}
    \caption{Behavior of the cost function when using the parametrization illustrated in Fig. \ref{fig:models} A. The \textit{Model A} represents the behavior of the cost function obtained by employing the HQCNN model based on the standard method, as illustrated in Fig. \ref{fig:modeloHibrido} A. On the other hand, the \textit{New Model} corresponds to the behavior of the cost function when using the HQCNN model obtained through the new method, as illustrated in Fig. \ref{fig:modeloHibrido} B.}
    \label{fig:loss_model_1}
\end{figure}

\begin{figure}[H]
    \centering
    \includegraphics[width=1\linewidth]{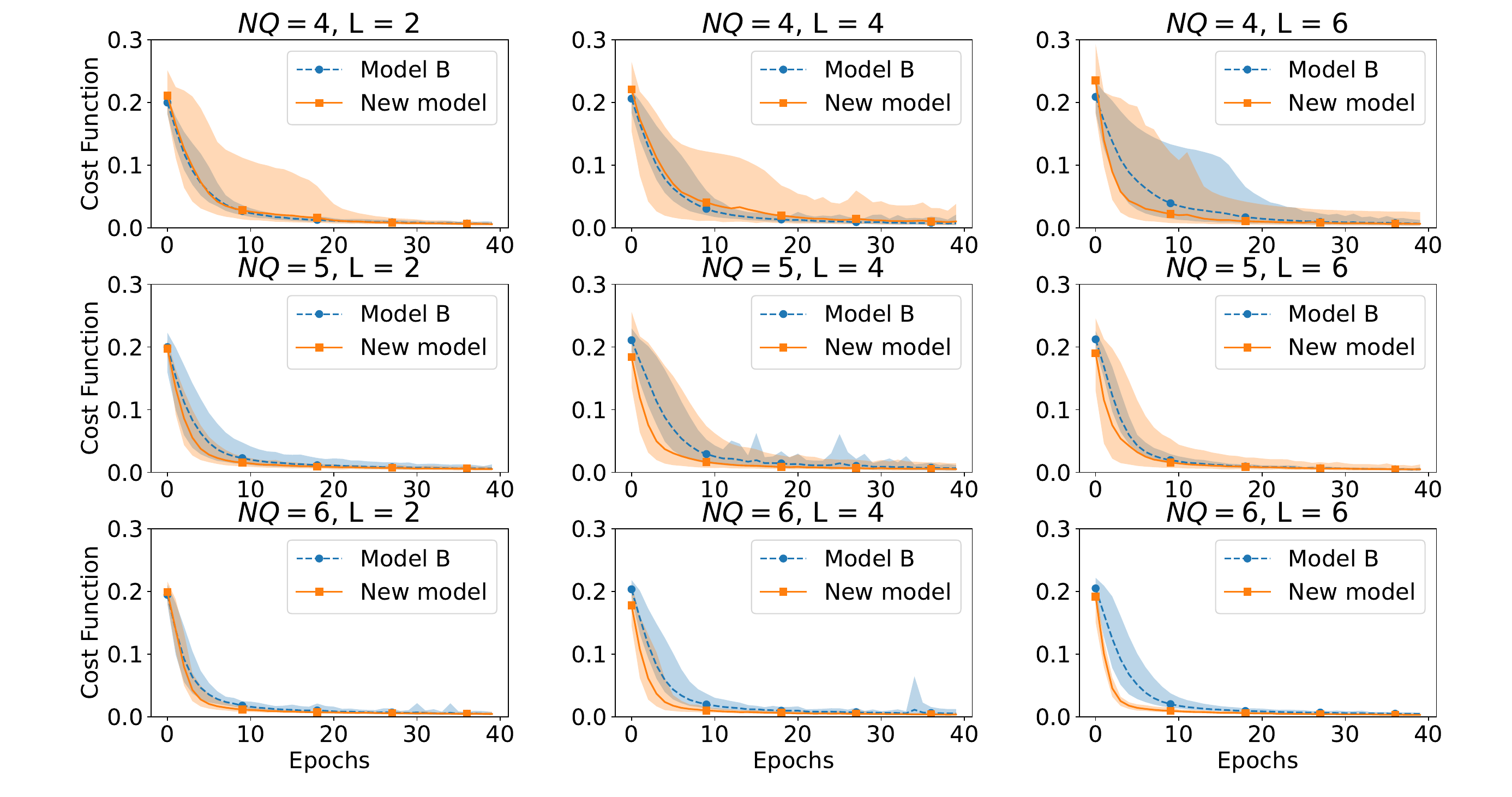}
    \caption{Behavior of the cost function when using the standard model, illustrated in Fig.~\ref{fig:modeloHibrido} A and referred to as \textit{Model B}, and the model based on the new method, illustrated in Fig.~\ref{fig:modeloHibrido} B and referred to as \textit{New Model}, employing the parametrization presented in Fig.~\ref{fig:models} B.}
    \label{fig:loss_model_2}
\end{figure}

\begin{figure}[H]
    \centering
    \includegraphics[width=1\linewidth]{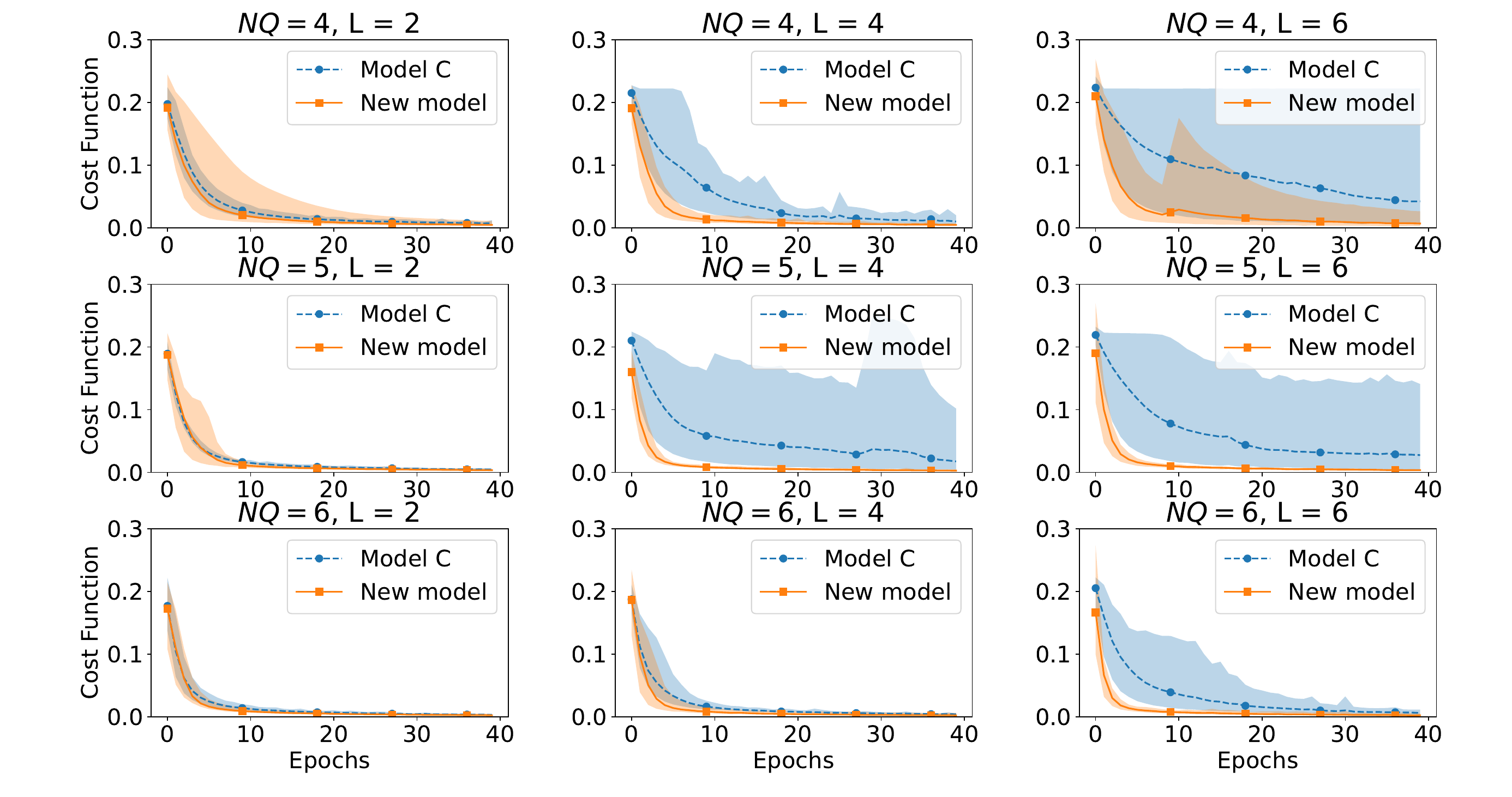}
    \caption{Comparative analysis of the behavior of the cost function between the standard model, illustrated in Fig.~\ref{fig:modeloHibrido} A and referred to as \textit{Model C}, and the model based on the new method, illustrated in Fig.~\ref{fig:modeloHibrido} B and referred to as \textit{New Model}, when using the parametrization presented in Fig.~\ref{fig:models} C.}
    \label{fig:loss_model_3}
\end{figure}

In this section, we present a comparison between the behavior of the cost function obtained using the standard HQCNN model and the behavior obtained with the HQCNN model constructed from the new method. As can be observed, when employing parametrizations A and B, illustrated in Fig.~\ref{fig:models} A and Fig.~\ref{fig:models} B, respectively, the results obtained in the analyzed cases are similar, as shown in Fig.~\ref{fig:loss_model_1} and Fig.~\ref{fig:loss_model_2}, respectively. However, when using parametrization C, presented in Fig.~\ref{fig:models} C, we can observe in Fig.~\ref{fig:loss_model_3} that the behavior of the cost function, when using the new method, shows a clear advantage over the standard method in some cases. Additionally, we can confirm that, for all three parametrizations, depending on the number of qubits $NQ$ and the depth $L$ used, the initialization of the parameters has a greater or lesser influence on both HQCNN models.

\section{Accuracy}\label{SI:Accuracy}

\begin{figure}[H]
    \centering
    \includegraphics[width=1\linewidth]{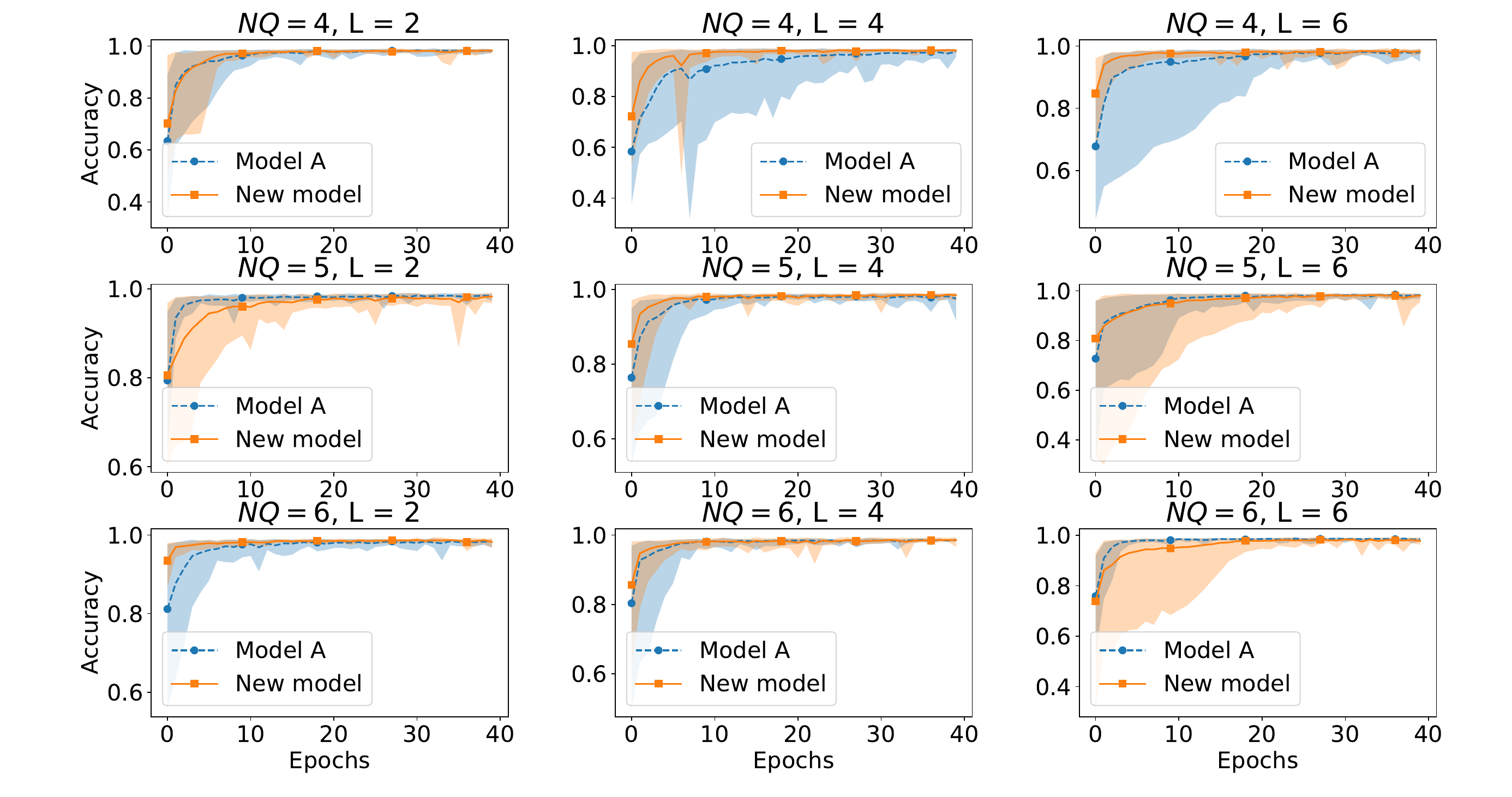}
    \caption{Accuracy behavior with respect to the test data obtained by the standard HQCNN model, using the parametrization illustrated in Fig.~\ref{fig:models} A and referred to as \textit{Model A}, and by the HQCNN model based on the new method, with the same parametrization, referred to as \textit{New Model}. In all cases, it is observed that the final results obtained by both models are similar, and that the initialization of the parameters had greater or lesser influence depending on the number of qubits and the depth $L$ used.}
    \label{fig:acc_model_1}
\end{figure}

\begin{figure}[H]
    \centering
    \includegraphics[width=1\linewidth]{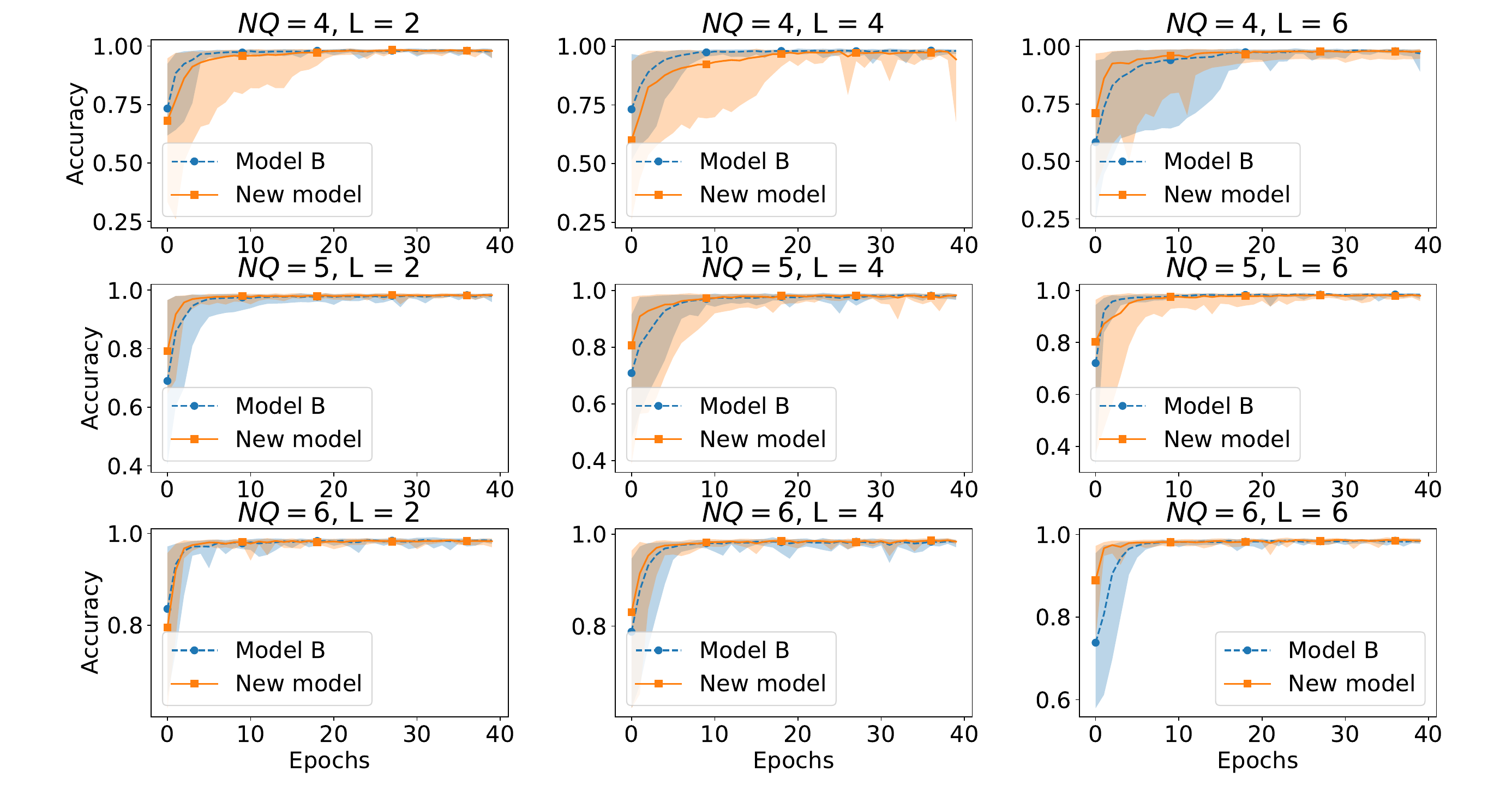}
    \caption{Accuracy behavior during training for the HQCNN models obtained using the standard method and the new method, with parametrization analogous to the one illustrated in Fig.~\ref{fig:models} B, referred to as \textit{Model B} and \textit{New Model}, respectively. In all cases, it is observed that the final results obtained by both models are similar, and the initialization of the parameters depended, to a greater or lesser extent, on the choice of hyperparameters, the number of qubits, and the depth $L$.
}
    \label{fig:acc_model_2}
\end{figure}

\begin{figure}[H]
    \centering
    \includegraphics[width=1\linewidth]{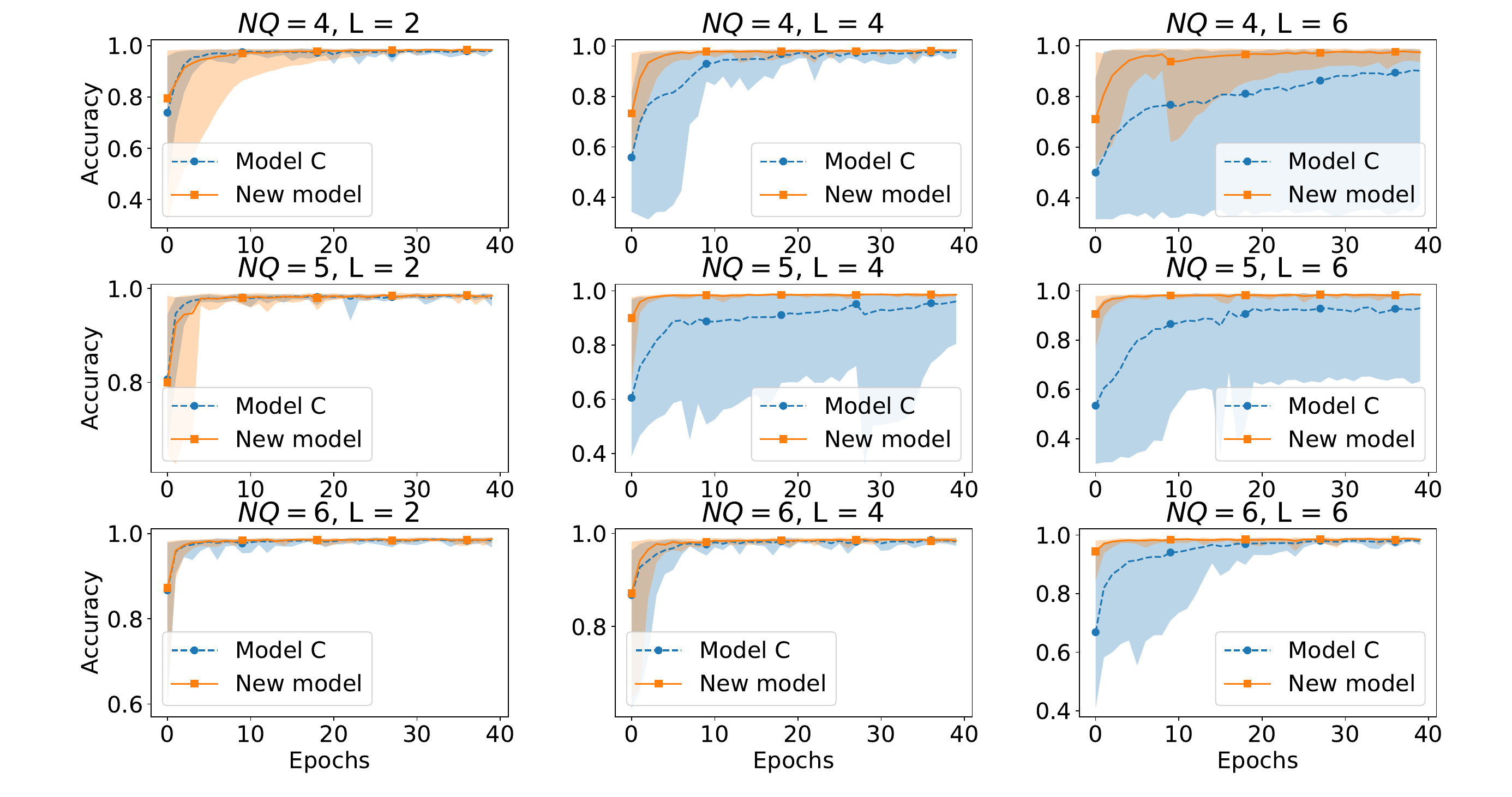}
    \caption{Accuracy obtained on test data during training for the standard HQCNN model, referred to as \textit{Model C}, and the HQCNN model using the new method, referred to as \textit{New Method}. It is observed that, in some cases, the accuracy obtained with the new method is superior to that obtained with the standard method. Additionally, the influence of parameter initialization depends on the choice of hyperparameters, specifically the number of qubits $NQ$ and the depth $L$.}
    \label{fig:acc_model_3}
\end{figure}

In this section, the accuracy behaviors during the training process with respect to the test data are presented for the HQCNN models using the standard method and the new method. As in the case of the cost function, it can be observed that when employing parametrizations A and B, illustrated in Fig.~\ref{fig:models} A and Fig.~\ref{fig:models} B, respectively, the results obtained in the analyzed cases are similar, as shown in Fig.~\ref{fig:acc_model_1} and Fig.~\ref{fig:acc_model_2}, respectively. Moreover, it can be observed that the accuracy obtained by the HQCNN model using the new method and parametrization C, presented in Fig.~\ref{fig:acc_model_3}, shows a clear advantage over the standard method in some cases. Finally, it is possible to observe that, for all three parametrizations, depending on the number of qubits $NQ$ and the depth $L$ used, the initialization of the parameters has a greater or lesser influence on both HQCNN models.

\end{document}